\documentclass[prb,twocolumn,showpacs]{revtex4}
\usepackage{amsfonts}
\usepackage{amsmath}
\usepackage{amssymb}
\usepackage{graphicx}

\providecommand{\U}[1]{\protect\rule{.1in}{.1in}}

\begin{document}

\title{Unconventional superconducting gap via spin fluctuations in
iron-vacancy ordered A$_{y}$Fe$_{2-x}$Se$_{2}$}
\author{Shin-Ming Huang$^{1}$ and Chung-Yu Mou$^{1,2,3}$}
\affiliation{$^{1}$Department of Physics, National Tsing Hua University, Hsinchu 30043,
Taiwan}
\affiliation{$^{2}$Institute of Physics, Academia Sinica, Nankang, Taiwan}
\affiliation{$^{3}$Physics Division, National Center for Theoretical Sciences, P.O.Box
2-131, Hsinchu, Taiwan}

\begin{abstract}
Based on an effective 12-orbital tight-binding model, we examine the
superconducting states induced by the antiferromagnetic fluctuations for
iron-vacancy-ordered A$_{y}$Fe$_{2-x}$Se$_{2}$. It is shown that due to the
broken reflection symmetry induced by the iron vacancies, new superconducting states
with $C_{4h}$ symmetry emerge. In particular, we show that in the  $C_{4h}$ symmetry,
symmetric axes of the pairing momenta do not need to coincide
with axes of the unit cell. As a result, in addition to the magnitude of
the pairing gap, the relative orientation of  the pairing wave function to the
lattice forms another degree of freedom for characterizing the
superconducting gap and can further help in gaining the
condensation energy. Nonetheless, similar to other iron-based
superconductors, the singlet ground state is still dominated by \textit{s}
-wave or \textit{d}-wave, which are nearly degenerate with anisotropic gaps.
Furthermore, \textit{s}-wave and \textit{d}-wave superconducting states are
separated by a quantum critical point controlled by the Hund's rule coupling
$J_{H}$.
\end{abstract}

\maketitle

\section{Introduction}

Since the discovery of high-temperature superconductivity in cuprates, it
has been a key interest to find the mechanism that causes the high $T_{c}$
and its unconventionality. After more than 20 years of intensive
investigations, however, the origin of high-$T_{c}$ superconductivity in
cuprates is still unsolved. The discovery of relatively high $T_{c}$
in iron-based superconductors \cite{Stewart2011,Oh2011} opens a
different route for high-temperature superconductivity and thus provides a
unique opportunity to re-clarify the physics of the unconventional
superconductivity. One of the features for the unconventional
superconductivity is the appearance of the sign-switched gap function. The
presence of disconnected Fermi surface (FS) sheets in iron-based
superconductors is an ingenious condition for the existence of the sign-switched order
parameters such as $s_{\pm }$-wave and \textit{d}-wave \cite
{Kuroki2008}. It is widely believed that the superconducting (SC) pairing is
driven by the inter-FS-sheet scattering or equivalently by the
antiferromagnetic (AFM) fluctuations \cite{Mazin2009,Zhai2009,Graser2009}
as the inter-FS-sheet scattering also leads to a strong AFM correlation.

The recent discovery of alkaline-intercalated iron-selenide superconductor A$
_{y}$Fe$_{2-x}$Se$_{2}$ with $T_{c}$ above 30K \cite{Gu2010} initiates an
interesting branch for exploring superconductivity in iron-based
superconductors. The chemically-stable structure A$_{0.8}$Fe$_{1.6}$Se$_{2}$
exhibits a $\sqrt{5}\times \sqrt{5}$ iron-vacancy order and
block-checkerboard antiferromagnetism \cite{Bao2011,Bacsa2011,Yan2011}. By
varying contents of iron or alkaline, the AFM state becomes unstable and the SC
state appears \cite{Fang2011,DMWang2011}. More recently, it is indicated that
the antiferromagnetic phase and the superconducting phase
may be separated in different space regions\cite{ZWang,Shen,Ryan2011,FChen2011,Ricci,Yuan}.
In particular, a scanning tunneling microscopy (STM) measurement \cite{Li2011}
shows that the sample would be spontaneously separated into the superconducting KFe$_{2}$Se$_{2}$
and the insulating K$_{x}$Fe$_{1.6}$Se$_{2}$.
Nevertheless, it still needs further clarification to see if the insulating region can be altered
into a metal by chemical doping or by applying pressure. First-principles
calculations indicate that by applying pressure to samples with the $\sqrt{5}\times \sqrt{5}$
iron-vacancy order, the system goes through two successive magnetic
transitions, from the semiconducting block-checkerboard AFM phase to the
metallic stripe (collinear) AFM phase and then to a metallic non-magnetic phase \cite{LChen2011}. These results agree with transport measurements and indicate that
a semiconductor-to-metal transition can be induced by the pressure\cite
{JGuo2011}. The issue of whether the iron-vacancy-ordered A$_{y}$Fe$_{2-x}$Se$_{2}$
supports superconductivity, however, remains unsettled as it is shown that as the system
became more metallic under pressure, $T_{c}$ gets smaller, and whether it suggests
the necessity of strong correlation for the emergence of superconductivity is an open question.

On the other hand, nuclear magnetic resonance (NMR) experiments find singlet superconductivity
and no pronounced spin fluctuations near $T_{c}$ \cite
{Yu2011,Torchetti2011,Ma2011}. The spin-lattice relaxation rate 1/\textit{T}$%
_{1}$\textit{T} shows the absence of the Hebel-Slichter coherence peak and a power-law
behavior below $T_{c}$, indicating an unconventional SC gap which is very likely
gapless. Although salient spin fluctuations are not present in
NMR, the possibility of spin-fluctuation-driven superconductivity should not
be excluded since the information of spin fluctuations would be hidden in signals\cite{comment1}.
A recent Raman measurement on the two-magnon scattering
shows that the scattering rate grows as the temperature approaches $T_{c}$
and undergoes a sudden drop when the system enters the SC phase \cite
{AMZhang2011}. It thus supports that superconductivity arises from magnetic
fluctuations.

Theoretically, previous works \cite
{Maier2011,FWang2011,YZhou2011,Das2011,Mazin2011,CFang2011} on
superconductivity in A$_{y}$Fe$_{2-x}$Se$_{2}$ were based on the band
structure of KFe$_{2}$Se$_{2}$ \cite{Shein2010,Cao1}. However, the
iron-vacancy order has been proved to induce a big change in the FS shape \cite
{Yan2011,Cao2}, though it is not confirmed so far by the angle-resolved photoemission
spectroscopy (ARPES) \cite{YZhang2011,Mou2011,Qian2011,LZhao2011}.
Therefore, if superconductivity could happen in the iron-vacancy ordering
phase, it is crucial to examine the pairing symmetry and pairing mechanism.

In this work, we investigate superconducting instability in
iron-vacancy-ordered A$_{y}$Fe$_{2-x}$Se$_{2}$. Previously, based on the
12-orbital tight-binding model that fits the band structure of K$_{y}$Fe$
_{2-x}$Se$_{2}$, we have studied the magnetic phase in the generalized
Hubbard model and succeeded in explaining the block-checkerboard AFM
instability from the Stoner's theory \cite{Huang}. Here based on the same
tight-binding model and starting from a non-magnetic metallic phase, we examine the superconductivity by the fluctuation-exchange (FLEX) approach
\cite{Bickers,Takimoto} . For singlet superconductivity, we examine the
pairing coupling constants for \textit{s}-wave and \textit{d}-wave states
from the effective scattering matrix. We find that similar to other iron-pnictide
superconductors \cite{Graser2009}, \textit{s}-wave and
\textit{d}-wave are closely degenerate. Furthermore, the gap functions in both symmetries are
highly anisotropic and have nodes. In addition, since the iron-vacancy order
lowers the symmetry of A$_{y}$Fe$_{2-x}$Se$_{2}$ to the group $C_{4h}$, the
gap functions do not need to be reflection symmetric and this implies that
there are degrees of freedom in defining origins of angles for the pairing
momenta. These extra degrees of freedom for the gap functions can further
increase the gain of the condensation energy. In addition, we find that the
pairing symmetry is
controllable by tuning the Hund's rule coupling $J_{H}$ and the critical
value at about $J_{H}=0.2U$. The Hund's rule coupling manipulates the phase
transition or crossover in superconductivity, antiferromagnetism \cite{Huang}
, or even the metallic transport \cite{Haule2009}, reflecting the
substantial role played by the orbital-correlation in the multi-orbital electronic
systems.

\section{Theoretical Method}

\begin{figure}[tbp]
\begin{center}
\includegraphics[height=2.578in,width=3.6754in] {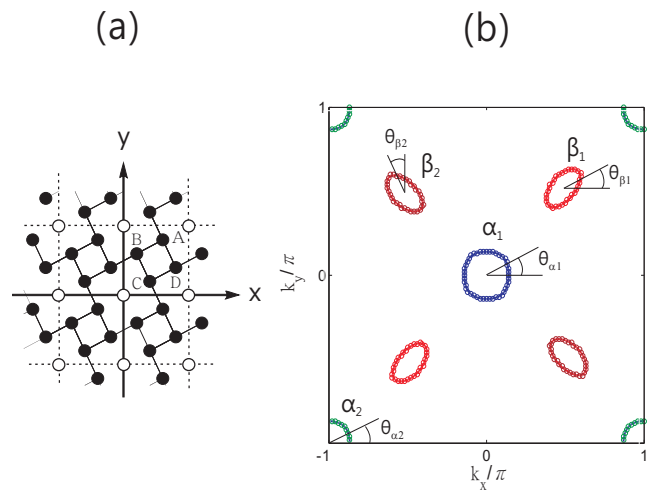}
\end{center}
\caption{(Color online) (a) The illustration of the Fe plane with the
left chiral $\protect\sqrt{5}\times \protect\sqrt{5}$ vacancy ordering. The
solid circles denote Fe atoms and the empty ones the vacancies. Every four
nearest vacancies enclose the unit cells. Each unit cell contains
four Fe atoms, labeled by \textit{A}, \textit{B}, \textit{C}, and \textit{D}. $x$ and $y$ are the primitive vectors. (b) The Fermi surface
results from the 12-orbital tight-binding model: two hole pockets ($\protect%
\alpha _{1}$ and $\protect\alpha _{2}$) and four\ electron pockets ($\protect%
\beta _{1}$ and $\protect\beta _{2}$ and their inverse partners). 216
\textit{k}-points on the Fermi surfaces are selected and shown by circles
(zoom in to see). Points on FS-$\protect\alpha _{1}$, $\protect\alpha _{2}$,
$\protect\beta _{1}$ and $\protect\beta _{2}$ are parameterized by the
angles, $\protect\theta _{\protect\alpha 1}$, $\protect\theta _{\protect%
\alpha 2}$, $\protect\theta _{\protect\beta 1}$, and $\protect\theta _{%
\protect\beta 2}$, respectively.}
\label{lattice}
\end{figure}

The vacancy ordered iron selenide A$_{y}$Fe$_{2-x}$Se$_{2}$ is a system with
one-fifth of Fe being taken off and forming a characteristic $\sqrt{5}\times%
\sqrt{5}$ pattern (Fig. \ref{lattice}(a)). The presence of the iron vacancy
changes the space group from \textit{I4/mmm} ($D_{4h}$) to \textit{I4/m} ($%
C_{4h}$). As a result, the four-fold rotational symmetry is retained without
the reflection symmetry with respect to the \textit{xz}, \textit{yz} and the
diagonal planes. The absence of reflection symmetry implies that there is no
symmetry axis. Hence there is an extra degree of freedom in defining the
zero value of angles for basis functions of $C_{4h}$. For instance, the
basis function $\cos (4\theta)$ becomes $cos [4(\theta-\theta_0)]$ with $%
\theta_0$ being an free parameter.

To implement the \textit{I4/m} symmetry for A$_{y}$Fe$_{2-x}$Se$_{2}$, we
have constructed a 12-orbital tight-binding model with four Fe atoms
(labeled by \textit{A}, \textit{B}, \textit{C}, and \textit{D}\textbf{)} per
cell and three $t_{2g}$ orbitals ($d_{xz}$, $d_{yz}$, and $d_{xy}$ with
x and y referring to  Fe-Fe directions) per Fe to
investigate the magnetic instability \cite{Huang}. The resulting FS is shown
in Fig. \ref{lattice}(b) with two hole pockets, $\alpha _{1}$ the one
centering on (0,0) and $\alpha _{2}$ on ($\pi $,$\pi $), and four electron
pockets, $\beta _{1}$ the one and its inversion around $\pm $($\pi /2$,$\pi
/2$) and $\beta _{2}$ around $\pm $($-\pi /2$,$\pi /2$). To calculate the SC
gap, we shall divide the momentum \textit{k} space into a lattice with 200$
\times $ 200 points and approximate states on the FS sheets by
picking totally 216 \textit{k} points with 44 points on FS-$\alpha _{1}$, 40
points on FS-$\alpha _{2}$, and 66 points both on FS-$\beta _{1}$ and $\beta
_{2}$. These \textit{k} points on FS sheets are characterized by angles, $%
\theta _{\alpha 1}$, $\theta _{\alpha 2}$, $\theta _{\beta 1}$, and $\theta
_{\beta 2}$ as illustrated in Fig. \ref{lattice}(b). All angles are relative
to the horizontal $k_{x}$ axis except that $\theta _{\beta 2}$ is relative
to the vertical $k_{y}$ axis due to that FS-$\beta _{2}$ is a 90${{}^{\circ }%
}$ rotation from FS-$\beta _{1}$. On the average, the energy resolution in the \textbf{k} space with 200$
\times $ 200 points within a 50meV window relative to the Fermi
energy is 0.6meV, while for the chosen 216 \textit{k} points, the momentum resolution is about $\Delta k=\pi /100$ and we
estimate the corresponding energy resolution is about 4meV.

We shall assume that the interaction between electrons on each Fe atom is
given by the generalized Hubbard model,
\begin{align}
H_{I}& =\underset{i}{{\textstyle\sum }}\underset{I}{{\textstyle\sum }}%
\left\{ U\underset{a=1}{{\textstyle\sum^{3}}}n_{aI,i\uparrow
}n_{aI,i\downarrow }\right.   \notag \\
& +\underset{a,b(a>b)}{{\textstyle\sum }}\left[ \left( U^{\prime }-\frac{%
J_{H}}{2}\right) n_{aI,i}n_{bI,i}-2J_{H}\mathbf{S}_{aI,i}\cdot \mathbf{S}%
_{bI,i}\right.   \notag \\
& \left. \left. +J_{C}\left( d_{aI,i\uparrow }^{\dag }d_{aI,i\downarrow
}^{\dag }d_{bI,i\downarrow }d_{bI,i\uparrow }+h.c.\right) \right] \right\} ,
\label{H_I}
\end{align}%
where \textit{I} is the index for four Fe atoms and \textit{a} is the index for three $t_{2g}$
orbitals. It was found \cite{Huang} that as the interaction turns on, the
spin fluctuation is strong at $\mathbf{q}=$($\pi $,$\pi $) and ($\pi $,$0$),
in which the former is related to the checkerboard antiferromagnetism and
the latter is related to the stripe antiferromagnetism. These two magnetic
states competes each other and the preference between them is controlled by
the Hund's rule coupling, $J_{H}$.

The strong spin fluctuation is a possible pairing mechanism. To investigate
the superconductivity arising from the exchange of spin and charge
fluctuations, we follow the FLEX approach \cite{Bickers,Takimoto}, in which
the effective singlet pairing scattering matrix is given by
\begin{align}
\Gamma _{cd}^{ab}(\mathbf{k,k}^{\prime };\omega )& =\left[ \frac{3}{2}\Gamma
_{s}\chi _{RPA}^{s}(\mathbf{k-k}^{\prime },\omega )\Gamma _{s}+\frac{1}{2}%
\Gamma _{s}\right.   \label{flex} \\
& \left. -\frac{1}{2}\Gamma _{c}\chi _{RPA}^{c}(\mathbf{k-k}^{\prime
},\omega )\Gamma _{c}+\frac{1}{2}\Gamma _{c}\right] _{cd}^{ab}\text{ },
\notag
\end{align}%
where $\omega$ is a real frequency and \textit{a, b, c}, and \textit{d} as before are orbital indices. In Eq.
(\ref{flex}), the spin and charge vertices $\Gamma _{s}$ and $\Gamma _{c}$
are $\Gamma _{s}^{\tau \tau ,\tau \tau }=U$, $\Gamma _{s}^{\tau \tau
^{\prime },\tau \tau ^{\prime }}=U^{\prime }$, $\Gamma _{s}^{\tau \tau ,\tau
^{\prime }\tau ^{\prime }}=J_{H}$, $\Gamma _{s}^{\tau \tau ^{\prime },\tau
^{\prime }\tau }=J_{C}$, and $\Gamma _{c}^{\tau \tau ,\tau \tau }=U$, $%
\Gamma _{c}^{\tau \tau ^{\prime },\tau \tau ^{\prime }}=-U^{\prime }+2J_{H}$%
, $\Gamma _{c}^{\tau \tau ,\tau ^{\prime }\tau ^{\prime }}=2U^{\prime }-J_{H}
$, $\Gamma _{c}^{\tau \tau ^{\prime },\tau ^{\prime }\tau }=J_{C}$, where
non-vanishing vertices are only between the same Fe, and $\tau $ labels
orbitals and $\tau \neq \tau ^{\prime }$. In the following, we will take the
relations $U^{\prime }=U-2J_{H}$ and $J_{C}=J_{H}$. The random phase
approximation (RPA) for spin and charge susceptibilities are given by
\begin{align}
\chi _{RPA}^{s}(\mathbf{q},\omega )& =\chi _{0}(\mathbf{q},\omega )\left[
1-\Gamma _{s}\chi _{0}(\mathbf{q},\omega )\right] ^{-1}, \\
\chi _{RPA}^{c}(\mathbf{q},\omega )& =\chi _{0}(\mathbf{q},\omega )\left[
1+\Gamma _{c}\chi _{0}(\mathbf{q},\omega )\right] ^{-1},
\end{align}%
with the bare susceptibility being given as
\begin{align}
\left[ \chi _{0}(\mathbf{q},\omega )\right] _{ab}^{cd}& =-\frac{1}{N}{\sum_{%
\mathbf{k},\mu ,\nu }}A_{c\mu }(\mathbf{k})A_{a\mu }^{\ast }(\mathbf{k}%
)A_{b\nu }(\mathbf{k+q})A_{d\nu }^{\ast }(\mathbf{k+q})  \notag \\
& \times \frac{n_{F}\left[ E_{\mu }(\mathbf{k})\right] -n_{F}\left[ E_{\nu }(%
\mathbf{k+q})\right] }{\omega +E_{\mu }(\mathbf{k})-E_{\nu }(\mathbf{k+q}%
)+i\delta },
\end{align}%
where $A_{a\mu }$ is the orbital-band transformation matrix from $\psi _{a}(%
\mathbf{k})=\underset{\mu }{{\textstyle\sum }}A_{a\mu }(\mathbf{k})\gamma
_{\mu }(\mathbf{k})$ \textbf{and }$n_{F}$ is the Fermi-Dirac distribution
function. In this work, we shall set the temperature to zero. Here $\chi _{0}(\mathbf{q}%
,\omega )$\ is calculated by using the bare Green's function not the dressed
one. The low energy physics of the scattering matrix is projected to
scattering among the FS sheets. We shall denote the scattering matrix with a
Cooper pair from $(\mathbf{k}^{\prime },-\mathbf{k}^{\prime })$ on FS-$\nu $
scattered to $(\mathbf{k},-\mathbf{k})$ on FS-$\mu $ by $\widetilde{\Gamma }%
_{\mu \nu }(\mathbf{k,k}^{\prime })$ \cite{comment2},
\begin{align}
\widetilde{\Gamma }_{\mu \nu }(\mathbf{k,k}^{\prime })& =\Re \left\{ {%
\sum\limits_{a,b,c,d}}A_{a\mu }^{\ast }(\mathbf{k})A_{d\mu }^{\ast }(-%
\mathbf{k})\right.   \label{gamma} \\
& \left. \times \Gamma _{cd}^{ab}(\mathbf{k,k}^{\prime };0)A_{b\nu }(\mathbf{%
k}^{\prime })A_{c\nu }(-\mathbf{k}^{\prime })\right\} .  \notag
\end{align}%
Since we consider the even-parity pairing, we shall symmetrize the scattering
matrix as $\widetilde{\Gamma }_{\mu \nu }^{even}(\mathbf{k,k}^{\prime })=%
\frac{1}{2}\left[ \widetilde{\Gamma }_{\mu \nu }(\mathbf{k,k}^{\prime })+%
\widetilde{\Gamma }_{\mu \nu }(\mathbf{k,-k}^{\prime })\right] $.
\begin{figure}[tbp]
\begin{center}
\includegraphics[
height=3.31in,
width=3.45in
]
{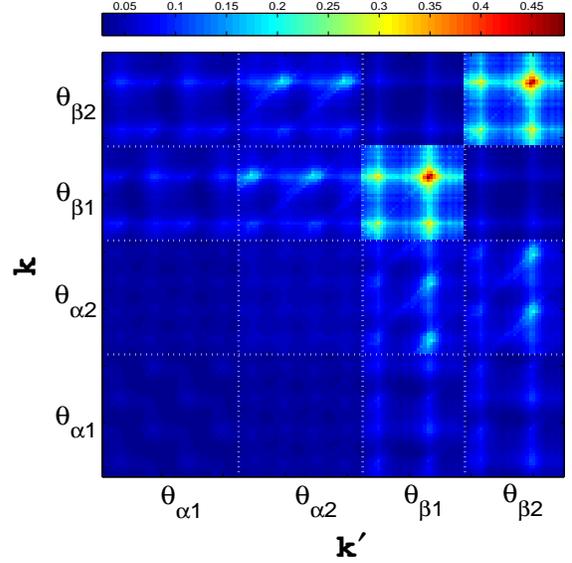}
\end{center}
\caption{(Color online) The weighted scattering matrix $\left[ v_{F}(\mathbf{
k})v_{F}(\mathbf{k}^{\prime })\right] ^{-1}\widetilde{\Gamma }_{\protect\mu
\protect\nu }^{even}(\mathbf{k,k}^{\prime })$ for $\mathbf{k,k}^{\prime }$
being parameterized by $\protect\theta _{\protect\alpha 1}$, $\protect\theta
_{\protect\alpha 2}$, $\protect\theta _{\protect\beta 1}$, and $\protect
\theta _{\protect\beta 2}$ as defined in Fig. \protect\ref{lattice}. Here the dotted
white lines separate different FS sheets. Values of $\theta$ range from 0${
{}^{\circ }}$ to 360${{}^{\circ }}$ in each block bounded by dotted white lines, which represents
the change of angles of  k as it moves around the corresponding Fermi surfaces.
Values of parameters are set with \textit{U}=1.0eV and $J_{H}$=0.2\textit{U}.}
\label{scattering_matrix}
\end{figure}

For a given gap function $g(\mathbf{k})$, a dimensionless coupling constant
is defined by \cite{scalapino1986}
\begin{equation}
\lambda=\frac{-\frac{1}{N_{p}}{\sum_{\mu,\nu}\sum_{\mathbf{k,k}^{\prime}}}%
\frac{1}{v_{F}(\mathbf{k})v_{F}(\mathbf{k}^{\prime})}g(\mathbf{k})\widetilde{%
\Gamma}_{\mu\nu}^{even}(\mathbf{k,k}^{\prime})g(\mathbf{k}^{\prime})}{{%
\sum_{\mu}\sum_{\mathbf{k}}}\frac{1}{v_{F}(\mathbf{k})}g^{2}(\mathbf{k})}.
\label{lambda}
\end{equation}
Here $\mathbf{k}\ $and $\mathbf{k}^{\prime}\ $are restricted within FS $\mu$
and $\nu$, respectively. $v_{F}(\mathbf{k})=\left\vert \nabla_{\mathbf{k}%
}E_{\mu}(\mathbf{k})\right\vert $ is the Fermi velocity, and $N_{p}$ (=216)
is the number of $\mathbf{k}$\ on Fermi surfaces we choose. The coupling
constant includes contributions from pocket-pocket scattering processes, and
it is helpful to extract them and to denote by a matrix $(\hat{\lambda}
)_{\mu\nu}$, satisfying $\lambda={\sum_{\mu,\nu}}(\hat{\lambda})_{\mu\nu}$,
where $\mu$, $\nu$ refer to FS sheets, $\mu$, $\nu$=1 to $\alpha_{1}$, 2 to $
\alpha_{2}$, 3 to $\beta_{1}$, and 4 to $\beta_{4}$.

The stationary condition in Eq. (\ref{lambda}) ($\delta\lambda\left[ g(
\mathbf{k})\right] /\delta g(\mathbf{k})=0$)\ leads to the eigenvalue
problem
\begin{equation}
-\frac{1}{N_{p}} {\textstyle\sum\limits_{\mathbf{k}^{\prime}}} \frac{1}{
v_{F}(\mathbf{k}^{\prime})}\widetilde{\Gamma}_{\mu\nu}^{even} (\mathbf{k,k}
^{\prime})g_{a}(\mathbf{k}^{\prime})=\lambda_{a}g_{a} (\mathbf{k}),
\label{gap_eqn}
\end{equation}
where the subscript $a$\ stands for different solutions. For a given
eigenvalue $\lambda_{a}$, its eigenfunction $g_{a}(\mathbf{k})$ determines
the symmetry of the gap, which could be \textit{s}-wave or \textit{d}-wave.
An \textit{s}-wave state is characterized by $g_{a}(\mathbf{k})=g_{a} (
\mathcal{R}\mathbf{k})$ and \textit{d}-wave by $g_{a}(\mathbf{k} )=-g_{a}(
\mathcal{R}\mathbf{k})$, where $\mathcal{R}$ is the 90${{}^\circ}$
-rotational operation on $\mathbf{k}$. Among the \textit{s}-wave solutions,
we pick up the largest one and set it to be $\lambda_{s}$ and its
eigenfunction to be $g_{s}$. Similarly, for the \textit{d}-wave, they are
denoted by $\lambda_{d}$ and $g_{d}$. The eigenvalue equation Eq. (\ref
{lambda}) is not identical to the Bethe-Salpeter equation in which magnitude of unity for
the eigenvalue stands for the formation of superconductivity. Instead,
eigenvalues in Eq. (\ref{lambda}) stand for the pairing strength. Therefore,
the SC state is determined by largest one of $\lambda_{s}$ and $\lambda_{d}$.

\section{Numerical Results}

\subsection{Scattering Matrix}

The scattering matrix shows some clues to the gap function. In order to gain
more condensation energy, the gap should be larger in the region where the
scattering matrix is maximal in magnitude. As the interaction is repulsive,
the gap becomes anisotropic and sign-changing. Fig. \ref{scattering_matrix}
shows the "weighted" scattering matrix $\left[ v_{F}(\mathbf{k})v_{F}(%
\mathbf{k}^{\prime })\right] ^{-1}\widetilde{\Gamma }_{\mu \nu }^{even}(%
\mathbf{k,k}^{\prime })$ for \textit{U}=1.0eV and $J_{H}$=0.2\textit{U}.
Here because the inverse Fermi velocity (IFV), $1/v_{F}(\mathbf{k})$, is
related to the density of states (DOS), it is used as a weighting factor at
each point. The scattering matrix contains intra and inter pocket scattering
channels and different pockets are discriminated by the white dotted lines.
The angles, $\theta _{\alpha 1}$, $\theta _{\alpha 2}$, $\theta _{\beta 1}$
and $\theta _{\beta 2}$ parameterizing momenta $\mathbf{k}$ ($\mathbf{k}%
^{\prime }$), increase from 0${{}^{\circ }}$ to 360${{}^{\circ }}$ as the
coordinate increases from the bottom (left) to the top (right). Although we
only show the result from one of the electron pocket $\beta _{1}$ ($\beta
_{2}$), due to the nature of being even parity for $\widetilde{\Gamma }_{\mu
\nu }^{even}(\mathbf{k,k}^{\prime })$, the result at the inversion point can
be deduced.
\begin{figure}[tbp]
\begin{center}
\includegraphics[
height=2.92in,
width=3.65in
]
{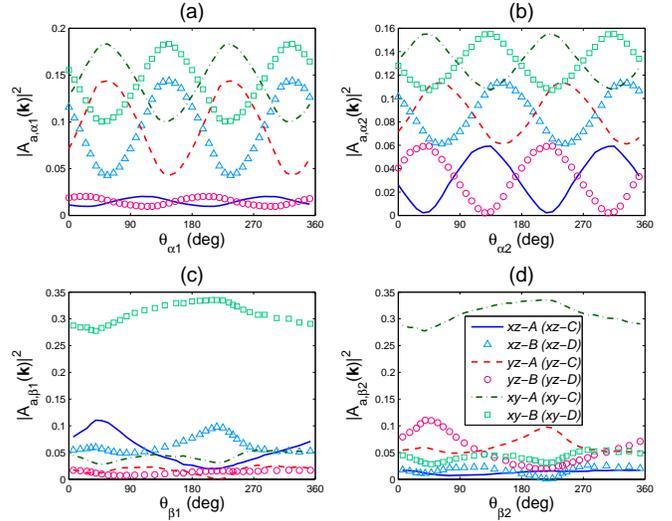}
\end{center}
\caption{(Color online) Orbital weight $\left\vert A_{a\protect\mu }(\mathbf{%
k})\right\vert ^{2}$ around FS sheets, $\protect\mu =\protect\alpha _{1}$
(a), $\protect\alpha _{2}$ (b), $\protect\beta _{1}$ (c), and $\protect\beta %
_{2}$ (d). Here lines (solid, dashed, and dash-dotted) represent Fe-\textit{A/C}
and legends (triangular, circular, and square) represent Fe-\textit{B/D}. Different
curves are related by rotational symmetry (see text).}
\label{orbital_weight}
\end{figure}

In Fig. \ref{scattering_matrix}, the intra-electron-pocket ($\beta _{1}$-$%
\beta _{1}$ or $\beta _{2}$-$\beta _{2}$) scattering is much stronger than
other channels and gives peaks at two momenta, $\theta _{\beta 1}\approx 48{%
{}^{\circ }}$ and $225{{}^{\circ }}$ as the antipodes (on the major axis) of
the electron pockets. The strong intensity of the scattering matrix at these
two momenta is expected because large IFV happens about the antipodes of the
elliptic FS ($1/v_{F}(\mathbf{k})$ is shown in Fig. \ref{s-wave}). One of
the reasons why intra-electron-pocket scattering is strong is due to the
much larger DOS on the electron pockets. This also explains why the
inter-electron-hole-pocket scattering is stronger than the intra- and the
inter-hole-pocket scattering. Furthermore, the intra-electron-pocket
scattering comes from exchange of fluctuations at vectors $\mathbf{q}\approx
$(0,0) and ($\pi $,$\pi $), and the latter is driven by the substantial
checkerboard antiferromagnetic fluctuations. The strength of DOS, however,
can not explain why the inter-electron-pocket ($\beta _{1}$-$\beta _{2}$)
scattering is much small because the stripe antiferromagnetic fluctuations at $
\mathbf{q}=$($\pi $,0) are not weak at all. A detailed analysis shows that the reduction of the inter-electron-pocket scattering is due
to the difference in the orbital-band matrix element. In Fig.
\ref{orbital_weight}, we show the orbital weight $\left\vert
A_{a\mu }(\mathbf{k)}\right\vert ^{2}$ for every FS sheet . Here lines represent Fe-\textit{A/C} and
legends with circles and triangles represent  Fe-\textit{B/D}.
Note that different curves exhibited in Fig.3 are related by rotational symmetry\cite{Huang}. For example, a 90${}^{\circ }$ rotation about the center of the unit cell
located at the position of Se changes Fe-A to Fe-B and $d_{xz}$ to $d_{yz}$. As a result,
as shown in Figs. \ref{orbital_weight}(a) or \ref{orbital_weight}(b), weight curve of
orbital-$(yz,B)$ is equal to that of orbital-$(xz,A)$ shifted by a phase $\pi /2$. Similarly, since orbital-$(xz,C)$ can be obtained by a 180${}^{\circ }$
rotation of orbital-$(xz,A)$, they have the same weight too. This is the relation
within the Fermi surface of $\alpha_1$ or $\alpha_2$. For relations between different Fermi surfaces,
since the same rotation changes FS-$\beta _{1}$ to FS-$\beta _{2}$ and relevant orbits also change under the rotation, the orbital components on different FS-$\beta $ would be
highly different. For instance, in Figs. \ref{orbital_weight}(c) and \ref{orbital_weight}(d),
the major component on FS-$\beta _{1}$ is orbital-$(xy,B)$
, while it is $(xy,A)$ on FS-$\beta _{2}$ and weights of
other orbits are much smaller. Because vertices in the fluctuation exchange
formula in Eq. (\ref{flex}) only couple orbitals on the same Fe, the
inter-electron-pocket scattering will be heavily reduced.  We have examined
the case of neglecting the orbital-band matrix elements in Eq. (\ref{gamma})
by setting $A_{a\mu }(\mathbf{k)}=1/\sqrt{12}$ and find that the
inter-electron-pocket scattering becomes comparable to the
intra-electron-pocket one. Therefore, the small inter-electron-pocket
scattering is due to the small magnitude of the orbital-band matrix element.
\begin{figure*}[tbp]
\centering{\includegraphics[
height=2.5002in,
width=5.6991in
]
{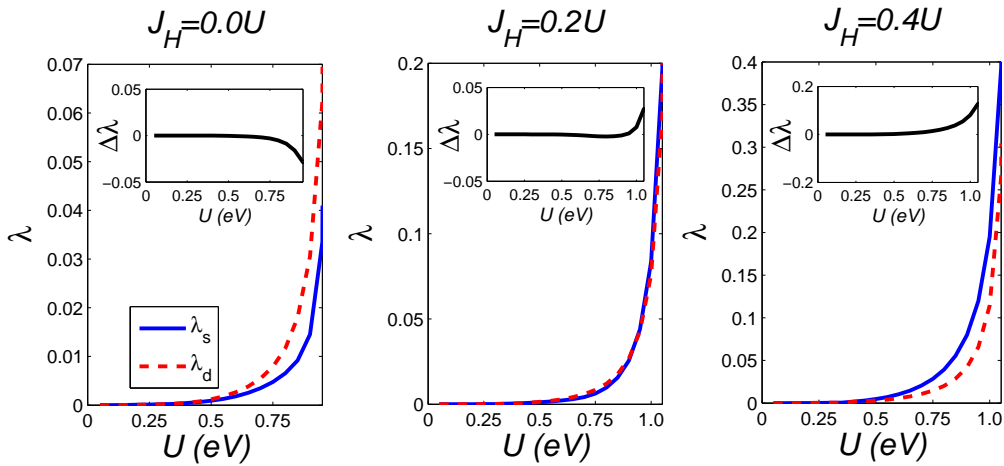} }
\caption{(Color online) Eigenvalues or dimensionless coupling constants $%
\protect\lambda _{s}$ (\textit{s}-wave) and $\protect\lambda _{d}$ (\textit{d%
}-wave) as functions of \textit{U} for $J_{H}$=0, 0.2\textit{U}, and 0.4%
\textit{U} from the left panel to the right one. Solid lines are for $
\protect\lambda _{s}$ and dashed lines are for $\protect\lambda _{d}$.
Insets are the difference of eigenvalues, $\Delta \protect\lambda =\protect%
\lambda _{s}-\protect\lambda _{d}$, to exhibit the favorable state. The
eigenvalues for $J_{H}$=0 are smaller, but they grow up fast and solutions
become unstable at about \textit{U}=1.0eV.}
\label{eigenvalues}
\end{figure*}

\subsection{Phase Diagram of Superconductivity}

After a better understanding of the scattering matrix, we now show the
results of eigenvalues from the eigenvalue problem in Eq. (\ref{gap_eqn}).
Fig. \ref{eigenvalues} displays the largest eigenvalues of \textit{s}-wave ($%
\lambda _{s}$) and \textit{d}-wave ($\lambda _{d}$) as functions of \textit{U%
} in three Hund's rule couplings, $J_{H}$=0, 0.2\textit{U} and 0.4\textit{U}%
. $\lambda _{s}$ is represented by solid blue lines while $\lambda _{d}$ is
represented by dashed red ones. As the interaction \textit{U} is increased,
quantum fluctuations enhances the SC pairing and then $\lambda $'s increase.
Moreover, we observe that at different $J_{H}$, the dimensionless coupling
constants for different symmetry grow in different speeds. As shown in the
insets where the differences of two eigenvalues $\Delta \lambda \equiv
\lambda _{s}-\lambda _{d}$ are plotted, the \textit{d}-wave is always more
stable against the \textit{s}-wave ($\Delta \lambda <0$) at $J_{H}$=0, while
the preference is reverse ($\Delta \lambda >0$) at $J_{H}$=0.4\textit{U}.
Between them at $J_{H}$=0.2\textit{U}, the \textit{d}-wave is favored for
\textit{U}$\lesssim $1.0eV but becomes unfavorable above that.
\begin{figure}[tbp]
\begin{center}
\includegraphics[
height=2.3324in,
width=3.1756in
]
{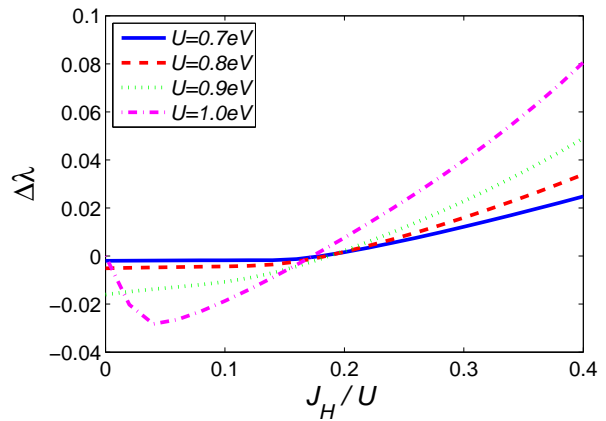}
\end{center}
\caption{(Color online) The eigenvalue differences between the \textit{s}
-wave and the \textit{d}-wave, $\Delta \protect\lambda $, as functions $%
J_{H} $/\textit{U} in four cases of \textit{U}: \textit{U}=0.7eV (solid line),
\textit{U}=0.8eV (dash line), \textit{U}=0.9eV (dotted line), and
\textit{U}=1.0eV (dash-dotted line). A critical $J_{H}$ ($\approx $0.2%
\textit{U}) is observed as the transition point between \textit{s}-wave and
\textit{d}-wave. }
\label{lambda_vs_jh}
\end{figure}

The above results indicate that the Hund's rule coupling is a parameter of
controlling the SC symmetry as well as the AFM order \cite{Huang}. This is
further supported by Fig. \ref{lambda_vs_jh}, in which $\Delta \lambda $
versus $J_{H}$ at different \textit{U} is plotted. Except for the small
phase space with \textit{U}=1.0eV and $J_{H}<0.4U$, $\Delta \lambda $
increases monotonously with $J_{H}$, indicating that the Hund's rule
coupling favors the \textit{s}-wave. However, $\Delta \lambda $ is not a
monotonic function of \textit{U}. There exists a critical $J_{H}$, $%
J_{H,c}\approx 0.2U$, separating two states; below it, $\Delta \lambda $
decreases as \textit{U} increases, but above it, $\Delta \lambda $ increases
with \textit{U}. Unless further symmetry is broken, \textit{s}-wave
and \textit{d}-wave do not mix and the phase transition between
them is first order.

\subsection{Gap Functions}

With the phase diagram of superconductivity, one still needs to know the gap
function explicitly to understand its physics, especially in this new type
of system with space group different from other iron-based superconductors.
We shall first examine the case of \textit{s}-wave solution in Fig. \ref%
{s-wave}\ and then the \textit{d}-wave one in Fig. \ref{d-wave}.

\subsubsection{\textit{s}-wave}

\begin{figure}[tbp]
\begin{center}
\includegraphics[
height=4.4884in,
width=3.1445in
]
{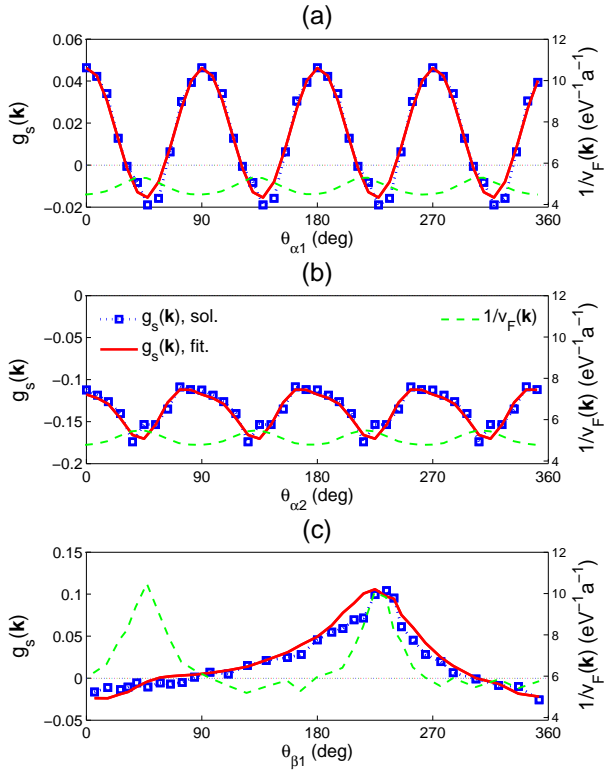}
\end{center}
\caption{(Color online) \textit{s}-wave eigenfunction $g_{s}(\mathbf{k})$
for \textit{U}=1.0eV and $J_{H}$=0.2\textit{U}. Panel (a) is along FS-$%
\protect\alpha _{1}$, (b) along FS-$\protect\alpha _{2}$, and (c) along FS-$%
\protect\beta _{1}$. Squares represent the eigenvalue-problem solution and
the solid lines are the fitting curves using function forms shown in the text.
Due to the \textit{s}-wave character, the gap function on FS-$
\protect\beta _{1}$\ and FS-$\protect\beta _{2}$\ is related by $g_{s}(%
\protect\theta _{\protect\beta 2})=g_{s}(\protect\theta _{\protect\beta 1})$
when $\protect\theta _{\protect\beta 2}=\protect\theta _{\protect\beta 1}$.
The factor, $1/v_{F}(\mathbf{k})$, being proportional to DOS, is also shown
by the dashed green lines. a is the lattice length and its value from
transmission electron microscopy \protect\cite{Wang2011} is about 6.15\AA .}
\label{s-wave}
\end{figure}

Following Ref.[\onlinecite{Graser2009}], we shall speak of \textit{s}-wave
by $g_{a}(\mathcal{R}\mathbf{k})=g_{a}(\mathbf{k})$ with $\mathcal{R}$ being
the 90${{}^{\circ }}$-rotational operation on $\mathbf{k}$. Fig. \ref{s-wave}
shows the \textit{s}-wave gap function $g_{s}(\mathbf{k})$ along\ FS sheets $%
\alpha _{1}$, $\alpha _{2}$, and $\beta _{1}$ for \textit{U}=1.0eV and $%
J_{H} $=0.2\textit{U}. $g_{s}(\mathbf{k})$ on FS-$\beta _{2}$ not shown is
the same as on FS-$\beta _{1}$ when $\theta _{\beta 2}=\theta _{\beta 1}$.
Blue squares are the eigenvalue-problem solution, while the solid red lines
are the fit guided by eye. One of the features is that such defined \textit{s%
}-wave gap have multi-nodes on FS-$\alpha _{1}$, $\beta _{1}$ and $\beta
_{2} $. Moreover, because there is no reflection symmetry, there is an extra
degree of freedom in defining origins of angles. As a result, the
fitting curves for \textit{s}-wave symmetry, chosen from lowest harmonic
functions, are found to be
\begin{eqnarray}
g_{s}(\theta _{\alpha 1}) &=&0.031\cos \left[ 4\left( \theta _{\alpha 1}-1{%
{}^{\circ }}\right) \right] +0.0153,  \notag \\
g_{s}(\theta _{\alpha 2}) &=&0.028\cos \left[ 4\left( \theta _{\alpha 2}+5{%
{}^{\circ }}\right) \right]  \notag \\
&+&0.008\cos \left[ 8\left( \theta _{\alpha 2}-22{{}^{\circ }}\right) \right]
-0.136,  \notag \\
g_{s}(\theta _{\beta 1}) &=&0.13\left[ \cos (k_{x}^{\prime })+\cos
(k_{y}^{\prime })\right]  \notag \\
&+&0.28\cos (k_{x}^{\prime \prime })\cos (k_{y}^{\prime \prime })-0.03\text{,
}  \label{swave}
\end{eqnarray}%
where $k_{x}^{\prime }=k_{x}\cos (32{{}^{\circ }})-k_{y}\sin (32{{}^{\circ }}
)$, $k_{y}^{\prime }=k_{x}\sin (32{{}^{\circ }})+k_{y}\cos (32{{}^{\circ }})$
and $k_{x}^{\prime \prime }=k_{x}\cos (4{{}^{\circ }})+k_{y}\sin (4{
{}^{\circ }})$, $k_{y}^{\prime \prime }=-k_{x}\sin (4{{}^{\circ }}
)+k_{y}\cos (4{{}^{\circ }})$. Here $\cos 4\theta $, $\cos 8\theta $, $\cos
(k_{x})+\cos (k_{y})$, and $\cos (k_{x})\cos (k_{y})$ are the common \textit{
s}-wave bases. (Because FS-$\beta _{1}$ is not located on symmetric point, a
different type of bases is adopted.) However, due to broken
reflection symmetry, origins of angles get shifted by $1{{}^{\circ }}$, $-5{
{}^{\circ }}$, $22{{}^{\circ }}$ and $20{{}^{\circ }}$. These angles are the
allowed degrees of freedom in the \ $C_{4h}$ symmetry and their precise
values are to minimize the total energy of the system.

Similar to the $s^{\pm }$-wave proposed in iron-pnictide superconductor \cite%
{Mazin2008}, where the SC order parameter on hole pockets and on electron
pockets have opposite signs to make superconductivity stable, here we have
the similar mechanism. In Fig. \ref{s-wave}, we see that the mean value of
the gap on FS-$\alpha _{2}$ has an opposite sign to those on FS-$\alpha _{1}$
and FS-$\beta $'s, which suggests these inter-pocket scatterings gain
energy. Due to their larger scattering strength, the $\alpha _{2}-\beta $'s
scattering processes gain most energy. The $\alpha _{1}-\beta $'s scattering
processes are not favorable due to large cancellation by the oscillating
behavior between positive and negative of gap on FS-$\alpha _{1}$. For the
intra-pocket scattering processes, because the Fermi surfaces are small and
if the fluctuations are smooth over the wave vector \textbf{q}, the
interaction would not able to change from repulsive to attractive. However,
the repulsive strength can be reduced by making the gap oscillatory and even
with the sign being changed. The reason why  gap functions take the
particular forms of Eq. (\ref{swave}) can be traced back to the behavior of
the inverse Fermi velocity (IFV), which is shown as the dashed lines in Fig. \ref{s-wave}. Clearly, the
oscillatory behavior of gap functions results from oscillatory IFV.
On FS-$\beta _{1}$, the IFV is peaked at $\theta _{\beta 1}\approx $225${
{}^{\circ }}$ where the gap is a peak too. The fact that the other point at $
\theta _{\beta 1}\approx $48${{}^{\circ }}$ with high IFV does not lead to a
large gap is because the repulsion can be lowered in the intra- and
inter-electron-pocket scattering processes.
Due to that gap functions on FS-$\beta$ change slowly, it is clear that
when the gap on FS-$\alpha _{2}$ is in-phase with IFV and dose not change sign,
one gains energy in $\alpha _{2}-\beta $\textbf's scatterings but loses
energy in the intra-pocket scatterings.
The compromise between the intra- and inter-pocket scattering processes
results in the gap function on FS-$\alpha _{1}$ with the gap function being about
$\frac{\pi }{2}$ out of phase relative to IFV. In the following, we list
numerical values of the coupling matrix $\hat{\lambda}_{s}/\lambda _{s}$,
which are consistent with the above explanations:
\begin{equation*}
\frac{\hat{\lambda}_{s}}{\lambda _{s}}=\left[
\begin{array}{cccc}
-0.0275 & 0.262\,8 & -0.103\,9 & -0.103\,9 \\
& -2.\,\allowbreak 829\,2 & 1.\,\allowbreak 603\,3 & 1.\,\allowbreak 603\,3
\\
&  & -1.\,\allowbreak 139\,8 & -0.194\,7 \\
&  &  & -1.\,\allowbreak 139\,8%
\end{array}%
\right] \allowbreak ,
\end{equation*}%
where due to that $\hat{\lambda}_{s}$ is a symmetric matrix, the
lower-triangular matrix elements are omitted for brevity, and $\lambda _{s}={%
\sum_{\mu ,\nu }}(\hat{\lambda}_{s})_{\mu \nu }=0.0837$.
\begin{figure}[tbp]
\begin{center}
\includegraphics[
height=4.4884in,
width=3.1445in
]
{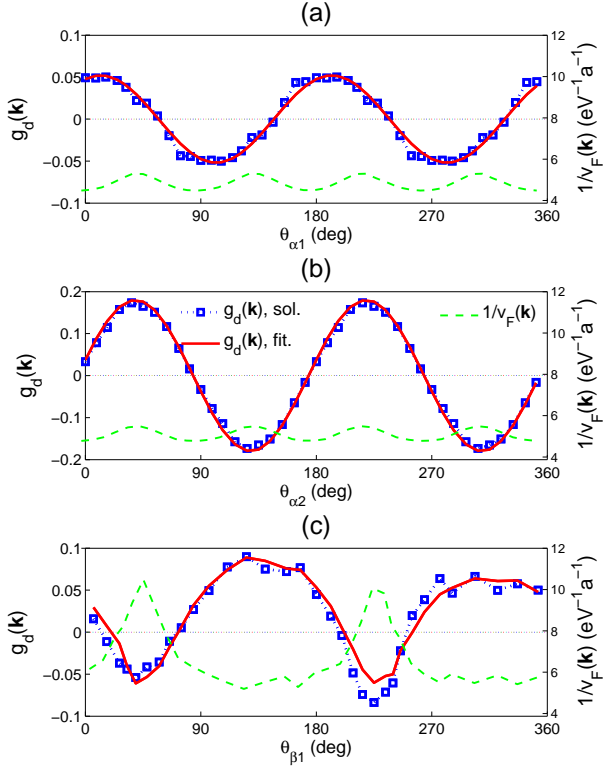}
\end{center}
\caption{(Color online) \textit{d}-wave eigenfunction $g_{d}(\mathbf{k})$
for \textit{U}=0.9eV and $J_{H}$=0.0. Due to the \textit{d}-wave character, $%
g_{d}(\protect\theta _{\protect\beta 2})=-g_{d}(\protect\theta _{\protect%
\beta 1})$ when $\protect\theta _{\protect\beta 2}=\protect\theta _{\protect%
\beta 1}$. Here legends represent the same meaning as those in Fig. \protect\ref
{s-wave} while the fitting curves are different and are discussed in the
text.}
\label{d-wave}
\end{figure}

\subsubsection{$\protect\bigskip$\textit{d}-wave}

Similar to the \textit{s}-wave, the \textit{d}-wave that we shall speak of
obeys $g_{a}(\mathcal{R}\mathbf{k})=-g_{a}(\mathbf{k})$ with $\mathcal{R}$
being the 90${{}^{\circ }}$-rotational operation on $\mathbf{k}$. In Fig. %
\ref{d-wave}, we show the \textit{d}-wave gap function $g_{d}(\mathbf{k})$
for \textit{U}=0.9eV and $J_{H}$=0. $g_{d}(\mathbf{k})$ on FS-$\beta _{1}$
and on FS-$\beta _{2}$ \textbf{is} equal in magnitude but has opposite
signs, \textit{i.e.,} $g_{d}(\theta _{\beta 2})=-g_{d}(\theta _{\beta 1})$
when $\theta _{\beta 2}=\theta _{\beta 1}$. It is seen that the gap
function of \textit{d}-wave always exhibits nodes on any Fermi surface.
The fitted functions in Fig. \ref{d-wave} (solid red) are given by
\begin{eqnarray}
g_{d}(\theta _{\alpha 1}) &=&0.052\cos \left[ 2\left( \theta _{\alpha 1}-12{%
{}^{\circ }}\right) \right] ,  \notag \\
g_{d}(\theta _{\alpha 2}) &=&0.18\cos \left[ 2\left( \theta _{\alpha 2}-39{%
{}^{\circ }}\right) \right] ,  \notag \\
g_{d}(\theta _{\beta 1}) &=&0.04\left[ \cos (k_{x}^{\prime })-\cos
(k_{y}^{\prime })\right]  \notag \\
&-&0.3\sin (2k_{x}^{\prime \prime })\sin (2k_{y}^{\prime \prime }),
\end{eqnarray}%
where $k_{x}^{\prime }=k_{x}\cos (20{{}^{\circ }})-k_{y}\sin (20{{}^{\circ }}%
)$, $k_{y}^{\prime }=k_{x}\sin (20{{}^{\circ }})+k_{y}\cos (20{{}^{\circ }})$%
, $k_{x}^{\prime \prime }=k_{x}\cos (2{{}^{\circ }})+k_{y}\sin (2{{}^{\circ }%
})$, and $k_{y}^{\prime \prime }=-k_{x}\sin (2{{}^{\circ }})+k_{y}\cos (2{%
{}^{\circ }})$. Note that instead of $\cos 4\theta $ (which is correct for the
\textit{s}-wave), \textit{d}-wave is represented by $\cos 2\theta $. Similar
to the \textit{s}-wave case, although nodes appear on every FS
sheet, they are shifted away from the diagonal directions.

The main cause for the occurrence of \textit{d}-wave differs from that for the \textit{s}-wave and
is primarily due to reduction of repulsion in
intra-pocket and inter-electron-pocket scatterings through its sign-changing character . As a result, both
antipodes on FS-$\beta _{1}$ ($\theta _{\beta 1}\approx $48${{}^{\circ }}$
and 225${{}^{\circ }}$) have large gaps in Fig. \ref{d-wave}, in comparison
to a single hump of the \textit{s}-wave in Fig. \ref{s-wave}. The
coupling constant matrix $(\hat{\lambda}_{d})_{\mu \nu }$ summarizes the
above effects
\begin{equation*}
\frac{\hat{\lambda}_{d}}{\lambda _{d}}=\left[
\begin{array}{cccc}
0.0033 & 0.0361 & 0.0066 & 0.0066 \\
& -0.0393 & 0.262\,3 & 0.262\,3 \\
&  & -0.0459 & -0.0066 \\
&  &  & -0.0459%
\end{array}%
\right] \allowbreak ,
\end{equation*}%
where $\lambda _{d}={\sum_{\mu ,\nu }}(\hat{\lambda}_{d})_{\mu \nu }=0.0305$
. Note that not only the repulsion between intra-pocket electrons is
reduced, due to the presence of nodes, the attraction between inter-pocket
electrons is also reduced. Although the \textit{d}-wave state might not gain
much energy from inter-pocket scattering processes relative to the \textit{s}
-wave state, it saves more energy from the reduction of intra-pocket
repulsion.
\begin{figure}[tbp]
\begin{center}
\includegraphics[
height=1.33in,
width=2.93in
]
{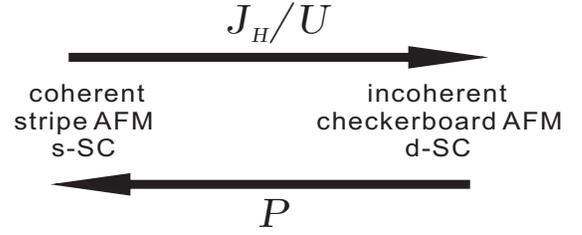}
\end{center}
\caption{Phase tendency of iron-vacancy-ordered A$_{y}$Fe$_{2-x}$Se$_{2}$  versus the Hund's rule
coupling over the Hubbard \textit{U} ($J_{H}/U$) and the pressure (\textit{P}
). Here arrows point to the direction that the corresponding parameters
increase their strength. Since increase of the pressure generally
reduces $J_{H}/U$ (see text), the increase of $
J_{H}/U $ or the decrease of \textit{P} would drive the multi-orbital electronic system
from coherent to incoherent, or from stripe to checkerboard AFM state, or
from \textit{d}-wave to \textit{s}-wave SC state found in this work. }
\label{phase_diagram}
\end{figure}

\section{Summary}

Proximity of antiferromagnetism and superconductivity in A$_{y}$Fe$_{1.6}$Se$
_{2}$ implies that the SC state could be derived from spin fluctuations.
Under this assumption and assuming that superconductivity exists in the $
\sqrt{5}\times \sqrt{5}$ iron-vacancy ordering state, we study the SC states
from the effective pairing interaction in the FLEX approximation based on
our previous 12-orbital tight-binding model \cite{Huang}. Similar to that of
iron-pnictide superconductors, for the spin-singlet superconductivity,
\textit{s}-wave and \textit{d}-wave states are found to be close in energy.
In particular, a quantum critical point is found at $J_{H}\approx 0.2U$,
below which the \textit{d}-wave prevails, while above which the \textit{s}
-wave wins over. Furthermore, unlike the iron-pnictide superconductors, the
iron-vacancy order lowers the symmetry to the group $C_{4h}$ so that the gap
functions do not need to be reflection symmetric. Therefore,
symmetric axes of the pairing momenta do not need to coincide
with axes of the unit cell. As a result, the relative orientations of the pairing wave functions to the
lattice, i.e., origins of angles for the pairing momenta, become new degrees of freedom for characterizing the superconducting gaps. This implies that the complete order parameters for
characterizing the superconductivity in A$_{y}$Fe$_{1.6}$Se$_{2}$ consist of both the
magnitude of the gap and the orientation of the pairing wave function relative to
the underlying lattice.

Finally, while our work focuses on the superconductivity, the finding of
quantum critical point controlled by $J_{H}$ in superconductivity is not an
accident. In the magnetism, we have shown that $J_{H}$ controls the magnetic
phase transition from the stripe to the checkerboard antiferromagnetism \cite
{Huang}. In addition, Haule and Kotliar \cite{Haule2009} show that the
crossover between coherence and incoherence is also determined by $J_{H}$.
Therefore, the Hund's rule coupling plays the dominant role in
determining localization, magnetism, and superconductivity of iron-based
superconductors.
Experimentally, it is observed that applying pressure drives
the magnetic phase transition\cite{LChen2011,JGuo2011} that accompanies the
metal-to-semiconductor transition.  Although in general, applying pressure affects itinerary electrons and could change $J_H$ and $\textit{U}$, the experimental results, when combined with our results on magnetism\cite{Huang}, indicate that
increase of the pressure generally reduces $J_{H}/U$ in this system. Therefore, the phase tendency
for iron-vacancy-ordered A$_{y}$Fe$_{2-x}$Se$_{2}$ can be summarized in Fig.\ref{phase_diagram} .
The tendency implies that by appropriately changing the applied pressure, transition between
different pairing symmetry of superconductivity can be induced.
While the phase tendency shown in Fig. \ref{phase_diagram} needs further supports
from experimental confirmation, our results provide a useful
guide in clarifying the origin of unconventional superconducting gaps and searching for a possible quantum phase transition in iron-vacancy-ordered A$_{y}$Fe$_{2-x}$Se$_{2}$.

\begin{acknowledgments}
We thank Dr. Ming-Chang Chung for useful discussions. This work was
supported by the National Science Council of Taiwan.
\end{acknowledgments}

\end{document}